# Eu$^{2+}$: a suitable substituent for Pb$^{2+}$ in CsPbX$_3$ perovskite nanocrystals?

*Firoz Alam,[1] K. David Wegner,[1] Stephanie Pouget,[2] Lucia Amidani,[3,4] Kristina Kvashnina,[3,4] Dmitry Aldakov,[1] Peter Reiss[1,*]*

[1] Univ. Grenoble Alpes, CEA, CNRS, IRIG, SyMMES, STEP, 38000 Grenoble, France

[2] Univ. Grenoble Alpes, CEA, IRIG, DEPHY, MEM-SGX, 38000 Grenoble, France

[3] The Rossendorf Beamline at ESRF – The European Synchrotron, CS40220, 38043 Grenoble Cedex 9, France

[4] Helmholtz Zentrum Dresden-Rossendorf (HZDR), Institute of Resource Ecology, PO Box 510119, 01314, Dresden

E-mail: peter.reiss@cea.fr

**ABSTRACT:** Europium is used to replace toxic lead in metal halide perovskite nanocrystals. They are synthesized by injecting cesium oleate in a solution of europium (II) bromide at an experimentally determined optimum temperature of 130°C. The obtained CsEuBr$_3$ NCs exhibit bright blue emission at 413 nm (FWHM: 30 nm) with a room temperature photoluminescence (PL) quantum yield of 39%. The emission originates from the Laporte allowed $4f^7 - 4f^65d^1$ transition of Eu$^{2+}$ and shows a PL decay time of 263 ns. Under optimized synthesis conditions long-term stability of the optical properties without any sign of oxidation to Eu$^{3+}$ is achieved, making the fully inorganic lead-free CsEuBr$_3$ NCs promising deep blue emitters for optoelectronics.

**INTRODUCTION**

Lead halide perovskites have not only become promising thin-film absorber materials in photovoltaics but they also inspire intense research efforts in form of colloidal semiconductor nanocrystals (NCs). Initially, organic-inorganic hybrid perovskite NCs such as methylammonium lead bromide (MAPbBr$_3$) were developed by Galian and Pérez-Prieto and later by Zhong and Rogach and coworkers, reaching up to unity photoluminescence quantum yield (PL QY) combined with narrow emission line widths.[1-5] These features make hybrid perovskite NCs very appealing for light-emitting applications, however, due to their high sensitivity to oxygen and moisture, efficient encapsulation strategies are required.[6-8] Fully inorganic lead halide perovskites with the formula ABX$_3$ (A=Cs$^+$, B=Pb$^{2+}$, X= Cl$^-$, Br$^-$ or I$^-$) have been already known since the end of the 19$^{th}$ century,[9] but their perovskite crystal structure and semiconducting nature were not reported until the 1950s.[10] They show an intrinsically higher stability than hybrid perovskites, albeit still lower than conventional II-VI, IV-VI or III-V semiconductor NCs due to their much higher ionic character.[11] While it turned out challenging to stabilize small-sized CsPbX$_3$ NCs in the strong quantum confinement regime below approx. 5 nm,[12] anion exchange has been shown to be an efficient way to fine-tune their optical and electronics properties, with band gap energies covering the entire visible range.[13-15] In the past few years, the high potential of CsPbX$_3$ NCs for use in diverse optoelectronic applications has been demonstrated, such as light emitting diodes,[16-18] solar cells[19] and photodetectors.[20-22]

Despite these appealing features the intrinsic toxicity of lead is a roadblock for practical applications of perovskite NCs, which triggered research on its replacement by less toxic metals. Most of these works focused on elements neighboring lead in the periodic table of elements, namely tin, bismuth and antimony.[23] In the case of tin, one has to consider the much higher oxidation sensitivity of Sn$^{2+}$ as compared to Pb$^{2+}$ whose divalent state is stabilized by relativistic effects (6$s^2$ inert electron pair). Jellicoe *et al.* first reported the successful synthesis of CsSnBr$_3$ and CsSnI$_3$ NCs, however, with low PL QYs (0.14% and 0.06%).[24] Trivalent Bi$^{3+}$ is isoelectronic with Pb$^{2+}$ and can form perovskite NCs of the formula A$_3$B$_2$X$_9$ crystallizing in the trigonal space group P-3m1.[25] The same type of structure can be adopted by Sb$^{3+}$. The highest reported PLQY for Cs$_3$Bi$_2$Br$_9$ QDs emitting at 410 nm (FWHM 48 nm) is 19.4%[26] and for Cs$_3$Sb$_2$Br$_9$ QDs emitting at the same wavelength (FWHM 41 nm) 46%.[27]

In contrast to these approaches, metal halide perovskite NCs involving rare earth (RE) ions for lead substitution have been essentially unexplored so far, although lanthanides have been used as dopants to modify the emission properties.[28-29] Here we report the first synthesis and main photophysical properties of colloidal CsEuBr$_3$ NCs. Europium has been chosen for its capacity of octahedral coordination in the divalent state and its almost identical ionic radius with Pb$^{2+}$ in this hexacoordinated configuration (117 pm / 119 pm). The obtained NCs exhibit a PL peak centered at 413 nm, a narrow emission line width (30 nm FWHM) and high PLQY of 39%, placing them among the brightest reported nanocrystalline emitters in the deep blue range.

**RESULTS AND DISCUSSION**

Alkali halide compounds of divalent rare earth (RE) ions are generally synthesized by high temperature reactions between stoichiometric amounts of the rare earth halide and the alkali halide.[30] In the case of europium (II) and cesium, the formation of perovskite-type ternary compounds of the formula CsEuX$_3$ (X=Cl, Br) was observed, while ARE$_2$X$_5$ compounds are obtained for smaller alkaline ions (Rb, K). In attempts to develop synthetic methods for nanoparticles of CsEuX$_3$ the low solubility and high oxidation sensitivity of the europium (II) halides used as starting reagents turned out to be the major challenges. As a starting point, the well-established hot-injection method reported for CsPbBr$_3$ NCs[11] was applied, using $EuBr_2$ instead of $PbBr_2$. In brief, a cesium oleate solution was quickly injected into a hot mixture of $EuBr_2$, oleic acid (OA) and oleylamine (OLA) in 1-octadecene (ODE). For an optimized reaction temperature of 130°C and reaction time of 60 s (vide infra) $CsEuBr_3$ NCs of approximately spherical shape and a mean size of 43 ± 7 nm were obtained as visible in the SEM and TEM images shown in **Figure 1**. Shorter reaction times (5-10 sec) do not give access to smaller NCs but yield ill-defined mixtures of larger particles and sheet-like structures (cf. **Fig. S1**). In the high-resolution TEM image (**Fig. 1b**) lattice planes can be unambiguously identified throughout the entire particle, confirming the high crystallinity of the obtained NCs. At the same time, they are sensitive to beam damage, as visible by the darker areas in the image. The lattice spacing of 0.295 nm corresponds to (004) planes in $CsEuBr_3$ (ICDD card #04-014-8774).[31] EDX analyses resulted in a Cs:Eu ratio of 1:0.92, i.e. close to the 1:1 ratio in $CsEuBr_3$, indicating that possible other phases like $Cs_2EuBr_4$ or $Cs_4EuBr_6$ do not form.[32] We note, however, a slightly elevated Br ratio of 3.86, which we attribute to remaining bromide after purification and/or oleylammonium bromide passivating the surface.

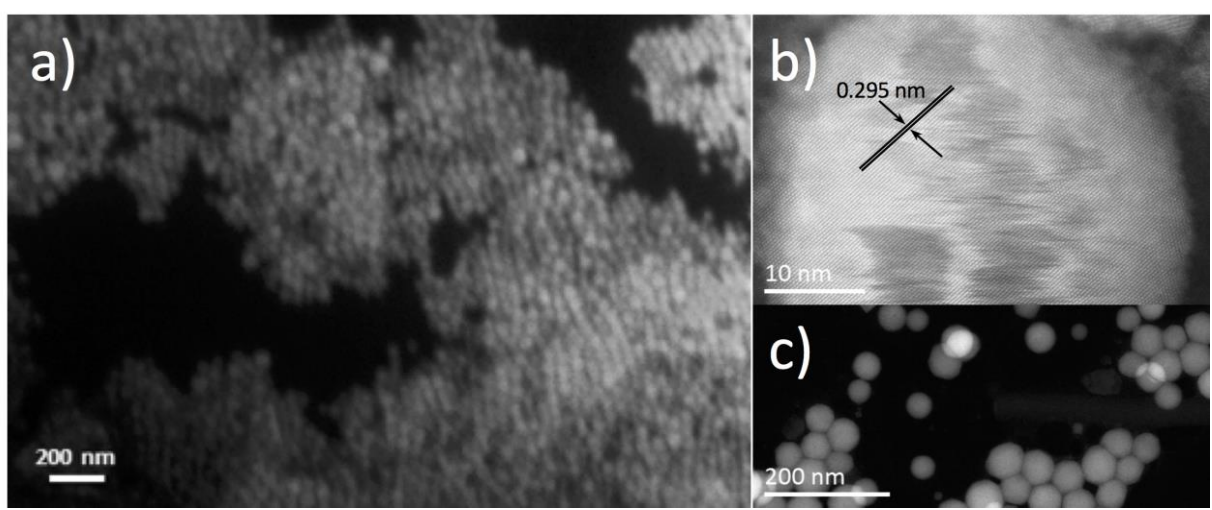

**Figure 2.** a) SEM image of the obtained $CsEuBr_3$ NCs (mean size 43 ± 7 nm). b) High-resolution, c) low-resolution STEM-HAADF images.

The structure of CsEuBr$_3$ single crystals has been reported to correspond to a distorted 3D perovskite structure, with an $a^-$ $a^-$ $c^+$ tilting scheme of the EuBr$_6$ octahedra.[31] It is isotypic to GdFeO$_3$ and crystallizes in the orthorhombic space group *Pbnm*. The powder X-ray data of the synthesized CsEuBr$_3$ NCs (**Fig. 2a**) reveals prominent peaks similar to the diffraction pattern of CsBr as well as a number of additional peaks characteristic fort the formation of the ternary structure. However, the latter do not fully match those observed in the powder data obtained from ground CsEuBr$_3$ single crystals.[33] Several reasons can be at the origin of these differences. In view of the high sensitivity against radiation damage observed in Eu$^{2+}$ doped CsBr used in X-ray radiography based on photostimulated luminescence (PSL),[34] we cannot fully exclude structural modifications occurring during the measurement, which could be at the origin of the observed differences. Furthermore, AREX$_3$ compounds are well-known to form several hettotypes arising from octahedral tilting and to frequently undergo phase transitions, which can result in the observed structural differences.[30]

X-ray absorption near edge structure (XANES) spectroscopy at the Eu L$_3$ edge has been used to identify the oxidation state of europium in the obtained NCs. XANES spectroscopy is one of the most powerful methods for the investigation of the electronic structure. Electrons in the X-ray absorption process are excited to unoccupied levels and give information about the chemical state. A chemical shift in the absorption edge is often assigned to a certain oxidation state. Better energy resolution of the XANES spectra can be obtained using the high energy resolution fluorescence detection (HERFD) mode, where a X-ray emission spectrometer is employed for data collection. **Fig. 2b** shows the HERFD XANES spectra at the Eu L$_3$ edge of CsEuBr$_3$ NCs compared to two reference systems, Eu$_2$O$_3$ and EuBr$_2$ with Eu(III) and Eu(II) oxidation states, respectively. The position of the main peak in the HERFD XANES spectrum of CsEuBr$_3$ NCs clearly demonstrates the presence of the Eu(II) oxidation state, based on the good correspondence with the EuBr$_2$ reference system. The exact contribution of the different chemical states in the Eu L$_3$ HERFD XANES data reported in **Fig. 2b** was estimated using the ITFA program.[35] The results (reported in SI) indicate that the spectrum of the CsEuBr$_3$ contains 99% of Eu(II) (with estimated root mean square error of less than 1%, *c.f.* SI).

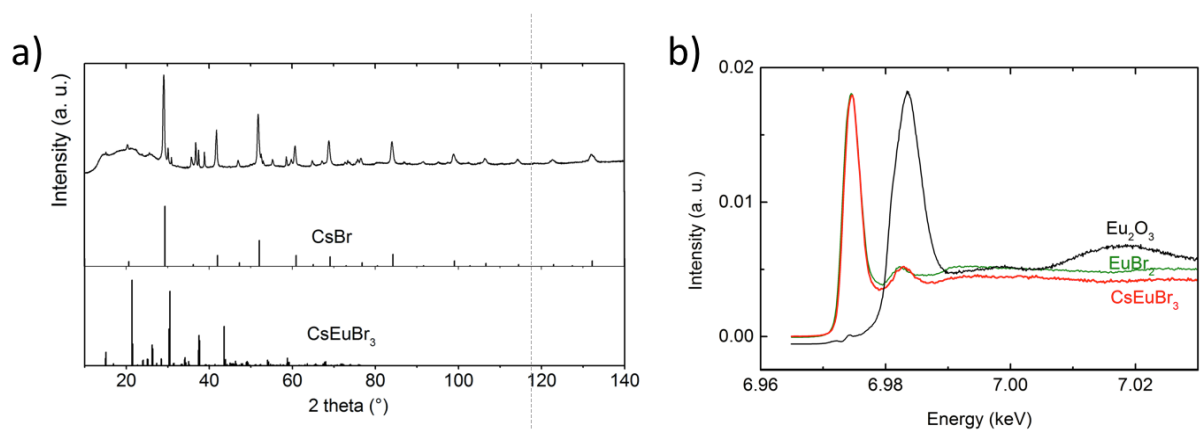

**Figure 2.** a) Powder X-ray diffractogram of CsEuBr$_3$ NCs. The broad feature in the 10-30° range originates from the two Kapton® foils used to protect the sample against oxidation during the measurement (cf. **Fig. S2**). For comparison, the diffraction patterns of bulk CsBr (ICCD #00-005-0588) and CsEuBr$_3$ (ICCD #04-014-8774) are also given. b) HERFD XANES spectrum at the Eu L$_3$ edge of CsEuBr$_3$ NCs compared with the reference systems Eu$_2$O$_3$ and EuBr$_2$.

An interesting feature of the Eu-based NCs is their strong photoluminescence (PL) in the deep blue range. In the PL spectrum of CsEuBr$_3$ NCs (**Fig. 3a**), an emission peak centered at 413 nm with a line width of 30 nm (FWHM) can be observed whose position is independent from the excitation wavelength. This peak is blue-shifted compared to Eu$^{2+}$-doped bulk CsBr or CsBr NCs (440 nm) and to CsEuBr$_3$ single crystals (450 nm).[29, 33, 36] In the former case, no perovskite structure is formed but $Eu^{2+}$ occupies cesium positions, which leads to the creation of $Cs^+$ vacancies in their immediate surrounding for charge compensation.[32] The observed bathochromic shift with respect to the single crystal data indicates differences within the coordination sphere of the $Eu^{2+}$ ion in $CsEuBr_3$ NCs, as already suggested by the X-ray data. For samples kept under inert atmosphere, no PL emission lines in the 600-700 nm range characteristic for the $5D_0 - 7F_J$ transitions of $Eu^{3+}$ ions are visible,[37] even when using longer excitation wavelengths. This is an indirect prove that no oxidation took place. Moreover, also after storing samples for more than three months in toluene or hexanes, no change in the PL spectra occurred (**Fig. S4**), demonstrating the long-term stability of $CsEuBr_3$ NCs. **Fig. 3c** shows the UV-vis absorption spectrum of the obtained $CsEuBr_3$ NCs in hexane. Upon strong dilution an absorption band peaking at 263 nm becomes visible, attributed to the $4f^7 - 4f^65d^1$ transition.[38] In contrast to the absorption spectra of $Eu^{2+}$ containing phosphors prepared in alkali halide melts,[39] no separation into two bands is observed. In particular, no second absorption band in the range of 300-350 nm due to the splitting of the $Eu^{2+}$ $5d$ orbitals by the crystal field is visible. In the 300-800 nm range the absorption spectrum is featureless (blue curve in **Fig. 3b**), while the excitation spectrum exhibits two small humps at around 320 and 355 nm (red curve).

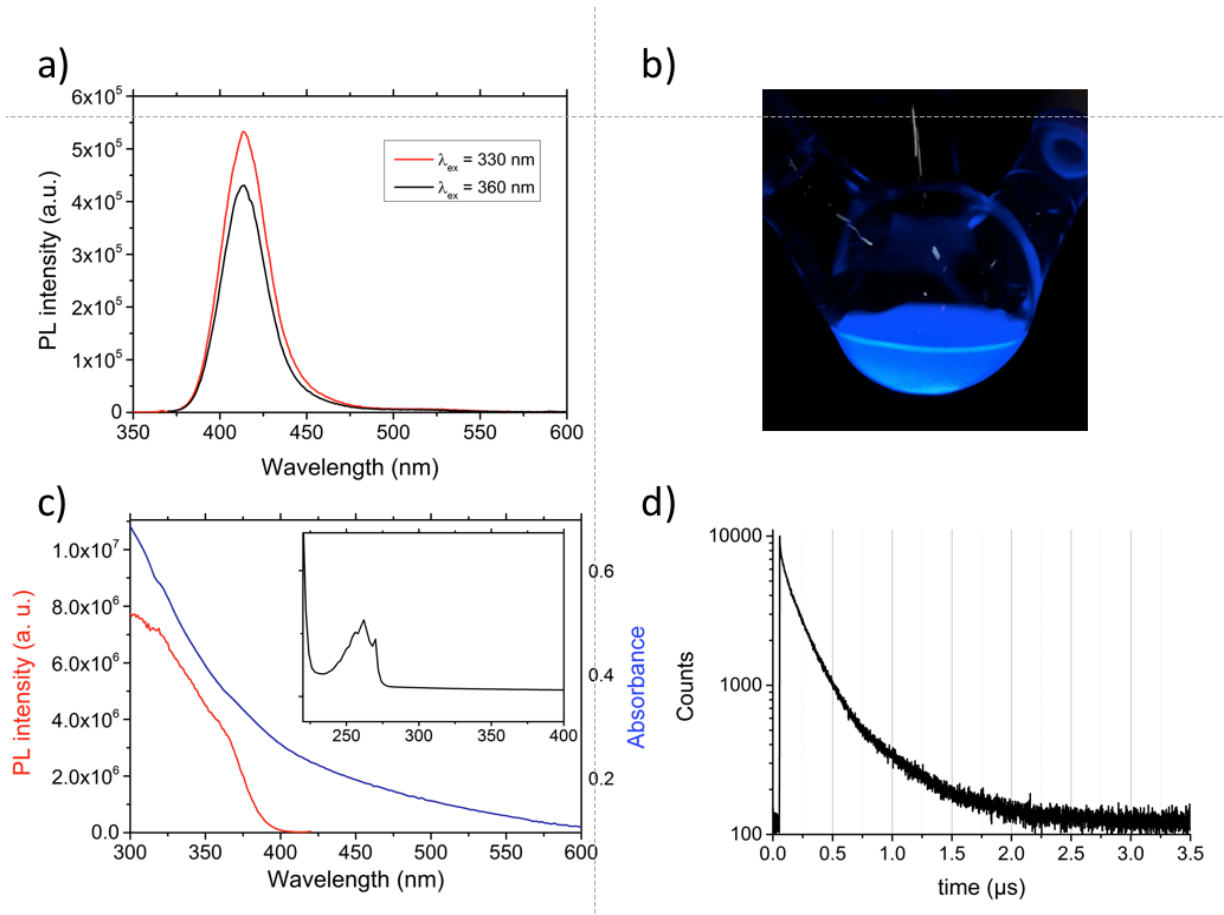

**Figure 3.** a) PL spectra of CsEuBr$_3$ NCs using two different excitation wavelengths in hexanes. The peak is centered at 413 nm (30 nm FWHM). b) Photograph of the colloidal solution under UV light (360 nm). c) UV-vis absorption spectrum (blue) and PLE spectrum (red, $\lambda_{em}$ = 425 nm). Insert: absorption spectrum of a strongly diluted colloidal solution. d) TRPL spectrum.

The PL decay curve obtained in time-resolved measurements (**Figure 1c**) was fitted with a triexponential function, resulting in lifetimes of $\tau_1$ 9.7 ns ($A_1$ 1.9), $\tau_2$ 124.5 ns ($A_2$ 35.7) and $\tau_3$ 296.0 ns ($A_3$ 62.5), yielding an average lifetime (intensity weighted) of 262.6 ns. As the underlying transition is Laporte-allowed, the lifetime is several orders of magnitude lower than that of well-known emitters involving Eu$^{3+}$ ions ($\tau$ in the millisecond range). The PL QY of the as-synthesized CsEuBr$_3$ NCs is 39% measured at room temperature using an integration sphere. Except for one example ($Cs_3Sb_2Br_9$ NCs emitting at 410 nm with a PLQY of 46%)[27] this value is higher than that of reported metal halide perovskite NCs luminescing in the deep blue range,[40] suggesting $CsEuBr_3$ NCs as a promising toxic heavy-metal-free emitter for optoelectronic applications.

While for lead-based perovskite NCs the variation of the reaction temperature is used to tune the NC size, here this parameter had a critical influence on the optical properties of the product. For too high temperatures either no emission was detectable (180°C) or partial oxidation of $Eu^{2+}$ took place (150°C), visible by the appearance of PL bands characteristic for $Eu^{3+}$ in the 700-800 nm range (cf. **Fig. S5**). As all reactions were carried out under strict air- and moisture-free conditions, oxidation could arise from side reactions of the used surface ligands, as for example amide formation[41] or ketonization,[42] leading to the *in situ* formation of minute amounts of water. Degradation of the optical properties continues to progress within a few days of storage. At too low reaction temperatures (e.g., 110°C) only partial conversion of the precursors and low reaction yields are observed, combined with limited stability (cf. **Fig. S6**). 130°C turned out to be an optimum value, combining high yield with long-term stability (> 3 months) of the optical properties (**Fig. S7**).

Additional experiments were conducted using $EuCl_2$ and $EuI_2$ as precursors. …

**CONCUSIONS**

$CsEuBr_3$ NCs synthesized using a hot injection method exhibit a narrow PL peak at 413 nm with a PLQY of 39%, which represents the highest reported value for lead-free $ABX_3$ perovskite NCs. No change in their optical properties was observed over several months when exposure to moisture was avoided. These features together with the absence of lead make CsEuBr$_3$ NCs promising candidates for optoelectronic applications. This first example of lanthanide halide perovskite NCs can also serve as a basis for the development of related systems involving the use of different halides and/or different lanthanide ions.

## EXPERIMENTAL SECTION

**Materials and Methods**

**Chemicals:** Cesium carbonate (Cs$_2$CO$_3$, Aldrich, 99.9%), oleic acid (OA, Fisher Chemicals, 70%), 1-octadecene (ODE, Sigma-Aldrich, 90%), oleylamine (OLA, Acros Organic, 80-90%), europium bromide (EuBr$_2$, Sigma-Aldrich, 99.99%), anhydrous toluene, hexane and acetonitrile (Sigma-Aldrich).

**Synthesis of Cs-oleate:** The synthesis was adapted from the method reported in ref [11]. 203.5 mg (0.62 mmol) of Cs$_2$CO$_3$, 10 mL of ODE and 0.625 mL of dried OA (1.97 mmol) were loaded into a 50 mL three-neck flask equipped with a condenser. The reaction mixture was continuously stirred and degassed for 1 h at 120°C under primary vacuum using a Schlenk line for the removal of oxygen and moisture. Later, the system was switched to argon atmosphere and the temperature was increased to 150 °C to get a clear solution of Cs-oleate. Cs-oleate solidifies at room temperature, it must be heated to 100 °C before injection.

**Synthesis of CsEuBr$_3$ NCs:** Colloidal CsEuBr$_3$ NCs were synthesized using a hot injection method. Typically, within a glove-box 58.6 mg (0.188 mmol) of europium bromide and 5 mL of ODE were loaded into a 50 mL three-neck flask. Outside the glove-box, the flask was connected to a condenser and degassed at 120 °C for 60 min using a Schlenk line. After backfilling with argon dried OLA and OA (0.5 mL each) were injected into the reaction mixture. Within 15 minutes, a clear colorless solution was obtained and the reaction temperature was increased to 130°C. The Cs-oleate solution (0.4 mL of the 0.12 M stock solution, preheated to 100 °C) was swiftly injected and 1 minute later the reaction was cooled down by immersing in an ice/water bath. After cooling to room temperature, the $CsEuBr_3$ NCs were purified by adding 1.5 mL of hexane followed by centrifugation at 1000 rpm for 5 minutes. The supernatant was discarded and a second cycle of purification was carried out by adding 1.5 mL of anhydrous acetonitrile to the precipitate followed by vortexing and centrifugation as before. Finally, the precipitated NCs were re-dispersed in a nonpolar solvent such as hexane or toluene for further analysis.

**Synthesis attempts for $CsEuCl_3$ and $CsEuI_3$ NCs:**

**Characterization:**

**Powder X-ray diffraction.** Powder X-ray diffraction was performed using a Panalytical X'Pert powder diffractometer equipped with a copper anode ($\lambda K_{\alpha 1}=$ 1.5406 Å, $\lambda K_{\alpha 2}=$ 1.5444 Å) and an X'Celerator 1D detector. It was configured in Bragg-Brentano geometry, with a variable divergence slit on the primary beam path and a set of antiscattering slits positioned before and after the sample. Axial divergence was limited by 0.02 rad Soller slits. The XRD samples were prepared in glove box by drop-casting a concentrated NCs dispersion in hexane on a disoriented silicon substrate and sealed with a double layer of airtight Kapton® foils within the sample holder.

**XANES measurements on ID26 at ESRF.** Eu $L_3$-edge XANES in High-Energy Resolution Fluorescence Detected (HERFD) mode were acquired on the ID26 beamline of the ESRF.[43] The incident energy was selected by a Si(311) double crystal monochromator. The footprint of the beam on the sample surface, oriented at 45 degrees to the incident beam direction, was 600 μm horizontal times 150 μm vertical. The HERFD XANES at Eu $L_3$ edge was collected using an X-ray emission spectrometer in Rowland geometry equipped with four spherically bent Ge(333) crystal analysers. The spectrometer was moved to the energy of the maximum of the Eu L$\alpha_1$ characteristic fluorescence line in order to collects only emitted photons in a 0.8 eV energy bandwidth around the maximum of the Eu L$\alpha_1$. Collecting the emitted photons on a bandwidth smaller than the core-hole lifetime broadening results in a sharpening of the XANES features compared to conventional fluorescence detected XANES, which integrates the full characteristic line.[44] The overall (incoming and emitted) energy resolution was 0.8 eV. All samples were measured in a liquid He cryostat kept at 20 K in order to minimize the X-ray beam damage and the contact with air.

We carefully checked for X-ray beam damage on all samples by acquiring fast XANES of the edge region. On sensitive samples, we adapted the thickness of Al attenuators and the scan time per spectrum to acquire HERFD XANES with minimal X-ray damage and we measured single XANES on several sample spots to have the desired statistics.

**SEM and EDX.** A ZEISS Ultra 55+ scanning electron microscope equipped with an EDX probe (acceleration tension, 20 keV; distance sample/ electron source, 7 nm) was used to obtain the images of the NCs and to determine the elemental composition. For sample preparation, a concentrated colloidal solution of $CsEuBr_3$ NCs in chloroform is drop-casted on a cleaned silicon substrate.

**Transmission Electron Microscopy.** Conventional transmission electron microscopy (TEM) images were acquired on a JEOL 3010 LaB6 microscopy equipped with a thermionic gun at 300 kV accelerating voltage. The samples were prepared by drop-casting diluted NC solution onto 200-mesh carbon-coated copper grids. STEM-HAADF images were recorded on an aberration corrected FEI Titan Themis$^3$ microscope using an acceleration voltage of 200 kV.

**UV-visible absorption spectroscopy.** UV-vis spectroscopy was performed with a Hewlett Packard 8452A single beam spectrophotometer operating in the wavelength range 190-820 nm and using a diode array for detection. The samples were prepared by diluting the NC dispersions in toluene or hexane in quartz cuvettes with a path length of 4 mm or 1 cm. The background was acquired using the pure solvent.

**Steady-state and time resolved photoluminescence.** Photoluminescence spectra were recorded using a modular Fluorolog FL3-22 system from Horiba-Jobin Yvon equipped with a double grating excitation monochromator and an iHR320 imaging spectrometer. A Hamamatsu R928P photomultiplier and quartz cuvettes with a path length of 1 cm were used for the measurement. For the PL measurements, the concentration of the samples was adjusted to an absorbance around 0.1 at the excitation wavelength.

Lifetimes of the NCs were obtained using a NanoLED pulsed source from Horiba (emission wavelength: 360 nm, repetition rate: 1 MHz). The output signal was controlled and analyzed with Data Station (v2.7) and Decay Analysis (v6.8) software from Horiba Scientist. The quantum yield measurements were performed at room temperature using an integration sphere, Hamamatsu Quantaurus-QY Absolute PL quantum yield spectrometer C11347-11.

## ACKNOWLEDGEMENTS

The authors acknowledge the French Research Agency ANR for financial support (Grants SuperSansPlomb ANR-15-CE05-0023-01 and PERSIL ANR-16-CE05-0019-02). KDW acknowledges the LABEX Serenade (Grant ANR 11-LABX-0064) for his post-doctoral funding. Hanako Okuna is thanked for TEM imaging. The authors greatfully acknowledge help of Tim Bohdan at the ID26 beamline of ESRF during the HERFD XANES measurements. L.A. and K.K. acknowledge support from the European Commission Council under ERC grant N759696.

**References**

[1] L. C. Schmidt, A. Pertegás, S. González-Carrero, O. Malinkiewicz, S. Agouram, G. Mínguez Espallargas, H. J. Bolink, R. E. Galian, J. Pérez-Prieto, *J. Am. Chem. Soc.* **2014**, *136*, 850-853.
[2] S. Gonzalez-Carrero, R. E. Galian, J. Perez-Prieto, *J. Mater. Chem. A* **2015**, *3*, 9187-9193.
[3] S. Gonzalez-Carrero, L. Francés-Soriano, M. González-Béjar, S. Agouram, R. E. Galian, J. Pérez-Prieto, *Small* **2016**, *12*, 5245-5250.
[4] F. Zhang, H. Zhong, C. Chen, X.-g. Wu, X. Hu, H. Huang, J. Han, B. Zou, Y. Dong, *ACS Nano* **2015**, *9*, 4533-4542.
[5] H. Huang, J. Raith, S. V. Kershaw, S. Kalytchuk, O. Tomanec, L. Jing, A. S. Susha, R. Zboril, A. L. Rogach, *Nature Communications* **2017**, *8*, 996.
[6] W. Deng, X. Xu, X. Zhang, Y. Zhang, X. Jin, L. Wang, S. T. Lee, J. Jie, *Adv. Funct. Mater.* **2016**, *26*, 4797-4802.
[7] Y. Ling, Z. Yuan, Y. Tian, X. Wang, J. C. Wang, Y. Xin, K. Hanson, B. Ma, H. Gao, *Adv. Mater.* **2016**, *28*, 305-311.
[8] H. Huang, M. I. Bodnarchuk, S. V. Kershaw, M. V. Kovalenko, A. L. Rogach, *ACS Energy Letters* **2017**, *2*, 2071-2083.
[9] H. L. Wells, *Zeitschrift für anorganische Chemie* **1893**, *3*, 195-210.
[10] C. K. Möller, *Nature* **1958**, *182*, 1436.
[11] L. Protesescu, S. Yakunin, M. I. Bodnarchuk, F. Krieg, R. Caputo, C. H. Hendon, R. X. Yang, A. Walsh, M. V. Kovalenko, *Nano Lett.* **2015**, *15*, 3692-3696.
[12] M. V. Kovalenko, L. Protesescu, M. I. Bodnarchuk, *Science* **2017**, *358*, 745-750.
[13] D. M. Jang, K. Park, D. H. Kim, J. Park, F. Shojaei, H. S. Kang, J.-P. Ahn, J. W. Lee, J. K. Song, *Nano Lett.* **2015**, *15*, 5191-5199.
[14] Q. A. Akkerman, V. D'Innocenzo, S. Accornero, A. Scarpellini, A. Petrozza, M. Prato, L. Manna, *J. Am. Chem. Soc.* **2015**.
[15] G. Nedelcu, L. Protesescu, S. Yakunin, M. I. Bodnarchuk, M. J. Grotevent, M. V. Kovalenko, *Nano Lett.* **2015**, *15*, 5635-5640.
[16] S. G. R. Bade, J. Li, X. Shan, Y. Ling, Y. Tian, T. Dilbeck, T. Besara, T. Geske, H. Gao, B. Ma, K. Hanson, T. Siegrist, C. Xu, Z. Yu, *ACS Nano* **2016**, *10*, 1795-1801.
[17] M. F. Aygüler, M. D. Weber, B. M. D. Puscher, D. D. Medina, P. Docampo, R. D. Costa, *J. Phys. Chem. C* **2015**, *119*, 12047-12054.
[18] S. Colella, M. Mazzeo, A. Rizzo, G. Gigli, A. Listorti, *J. Phys. Chem. Lett.* **2016**, *7*, 4322-4334.
[19] A. Swarnkar, A. R. Marshall, E. M. Sanehira, B. D. Chernomordik, D. T. Moore, J. A. Christians, T. Chakrabarti, J. M. Luther, *Science* **2016**, *354*, 92-95.
[20] P. Ramasamy, D.-H. Lim, B. Kim, S.-H. Lee, M.-S. Lee, J.-S. Lee, *Chem. Commun.* **2016**, *52*, 2067-2070.
[21] L. Lv, Y. Xu, H. Fang, W. Luo, F. Xu, L. Liu, B. Wang, X. Zhang, D. Yang, W. Hu, A. Dong, *Nanoscale* **2016**, *8*, 13589-13596.
[22] M. I. Saidaminov, M. A. Haque, M. Savoie, A. L. Abdelhady, N. Cho, I. Dursun, U. Buttner, E. Alarousu, T. Wu, O. M. Bakr, *Adv. Mater.* **2016**, *28*, 8144-8149.
[23] J. Sun, J. Yang, J. I. Lee, J. H. Cho, M. S. Kang, *J. Phys. Chem. Lett.* **2018**, *9*, 1573-1583.

[24]  T. C. Jellicoe, J. M. Richter, H. F. J. Glass, M. Tabachnyk, R. Brady, S. E. Dutton, A. Rao, R. H. Friend, D. Credgington, N. C. Greenham, M. L. Böhm, *J. Am. Chem. Soc.* **2016**, *138*, 2941-2944.
[25]  M. Leng, Z. Chen, Y. Yang, Z. Li, K. Zeng, K. Li, G. Niu, Y. He, Q. Zhou, J. Tang, *Angew. Chem. Int. Ed.* **2016**, *55*, 15012-15016.
[26]  M. Leng, Y. Yang, K. Zeng, Z. Chen, Z. Tan, S. Li, J. Li, B. Xu, D. Li, M. P. Hautzinger, Y. Fu, T. Zhai, L. Xu, G. Niu, S. Jin, J. Tang, *Adv. Funct. Mater.* **2018**, *28*, 1704446.
[27]  J. Zhang, Y. Yang, H. Deng, U. Farooq, X. Yang, J. Khan, J. Tang, H. Song, *ACS Nano* **2017**, *11*, 9294-9302.
[28]  G. Pan, X. Bai, D. Yang, X. Chen, P. Jing, S. Qu, L. Zhang, D. Zhou, J. Zhu, W. Xu, B. Dong, H. Song, *Nano Lett.* **2017**, *17*, 8005-8011.
[29]  Z. Yang, Z. Jiang, X. Liu, X. Zhou, J. Zhang, W. Li, *Advanced Optical Materials* **2019**, *7*, 1900108.
[30]  G. Meyer, *Prog. Solid State Chem.* **1982**, *14*, 141-219.
[31]  H. Ehrenberg, H. Fuess, S. Hesse, J. Zimmermann, H. von Seggern, M. Knapp, *Acta Crystallographica Section B* **2007**, *63*, 201-204.
[32]  P. Hackenschmied, G. Schierning, M. Batentschuk, A. Winnacker, *J. Appl. Phys.* **2003**, *93*, 5109-5112.
[33]  S. Hesse, J. Zimmermann, H. v. Seggern, H. Ehrenberg, H. Fuess, C. Fasel, R. Riedel, *J. Appl. Phys.* **2006**, *100*, 083506.
[34]  J. Zimmermann, S. Hesse, H. von Seggern, M. Fuchs, W. Knüpfer, *J. Lumin.* **2005**, *114*, 24-30.
[35]  A. Rossberg, T. Reich, G. Bernhard, *Analytical and bioanalytical chemistry* **2003**, *376*, 631-638.
[36]  P. Hackenschmied, G. Zeitler, M. Batentschuk, A. Winnacker, B. Schmitt, M. Fuchs, E. Hell, W. Knüpfer, *Nuclear Instruments and Methods in Physics Research Section B: Beam Interactions with Materials and Atoms* **2002**, *191*, 163-167.
[37]  U. T. D. Thuy, A. Maurice, N. Q. Liem, P. Reiss, *Dalton Transactions* **2013**.
[38]  K. E. Johnson, J. N. Sandoe, *J. Chem. Soc. (A)* **1969**, 1694-1697.
[39]  R. Reisfeld, A. Glasner, *J. Opt. Soc. Am.* **1964**, *54*, 331-333.
[40]  N. K. Kumawat, X.-K. Liu, D. Kabra, F. Gao, *Nanoscale* **2019**, *11*, 2109-2120.
[41]  M. Protiere, P. Reiss, *Chem. Commun.* **2007**, 2417-2419.
[42]  A. Cros-Gagneux, F. Delpech, C. Nayral, A. Cornejo, Y. Coppel, B. Chaudret, *J. Am. Chem. Soc.* **2010**, *132*, 18147-18157.
[43]  C. Gauthier, V. A. Solé, R. Signorato, J. Goulon, E. Moguiline, *Journal of Synchrotron Radiation* **1999**, *6*, 164-166.
[44]  P. Glatzel, T.-C. Weng, K. Kvashnina, J. Swarbrick, M. Sikora, E. Gallo, N. Smolentsev, R. A. Mori, *J. Electron. Spectrosc. Relat. Phenom.* **2013**, *188*, 17-25.